\documentstyle[aps,preprint]{revtex}
\def\lambar{{\mathchar'26\mkern-10mu\lambda}}
\begin{document}
\draft
\title { Extended Hauser-Feshbach Method for Statistical Binary-Decay of
Light-Mass Systems}

\author { T. Matsuse$^{*}$, C. Beck, R. Nouicer, D. Mahboub }

\address{\it Centre de Recherches Nucl\'eaires, Institut National de Physique
Nucl\'eaire et de Physique des Particules - Centre National de la Recherche
Scientifique/Universit\'e Louis Pasteur, B.P.28, F-67037 Strasbourg Cedex 2,
France }

\date{\today}
\maketitle

\begin{abstract}

An Extended Hauser-Feshbach Method (EHFM) is developed for {\bf \it light}
heavy-ion fusion reactions in order to provide a detailed analysis of all the
possible decay channels by including explicitly the fusion-fission phase-space
in the description of the cascade chain. The mass-asymmetric fission component
is considered as a complex-fragment binary-decay which can be treated in the
same way as the light-particle evaporation from the compound nucleus in
statistical-model calculations. The method of the phase-space integrations for
the binary-decay is an extension of the usual Hauser-Feshbach formalism to be
applied to the mass-symmetric fission part. The EHFM calculations include
ground-state binding energies and discrete levels in the low excitation-energy
regions which are essential for an accurate evaluation of the phase-space
integrations of the complex-fragment emission (fission). In the present
calculations, EHFM is applied to the first-chance binary-decay by assuming that
the second-chance fission decay is negligible. In a similar manner to the
description of the fusion-evaporation process, the usual cascade calculation of
light-particle emission from the highly excited complex fragments is applied.
This complete calculation is then defined as EHFM+CASCADE. Calculated
quantities such as charge-, mass- and kinetic-energy distributions are compared
with inclusive and/or exclusive data for the $^{32}$S+$^{24}$Mg and
$^{35}$Cl+$^{12}$C reactions which have been selected as typical examples.
Finally, the missing charge distributions extracted from exclusive measurements
are also successfully compared with the EHFM+CASCADE predictions.

\end{abstract}


\pacs{{\bf PACS} numbers: 25.70.Jj, 25.70.Gh, 25.70.Lm, 24.60.Dr}


\centerline{\bf 1 - INTRODUCTION}

\bigskip

For heavy-ion induced reactions in both the low- and intermediate-energy
regimes, the emission of complex fragments (or intermediate-mass fragments
IMF's) has been considered to be one of the most useful probes for the
investigation of the different reaction mechanisms involved in fission-like
phenomena for a wide mass-range of nuclear systems
\cite{So84,Sa86,Sh87,Au87,Ch88,Go91,Be92,Na92}. It has been shown that for
composite systems in the {\bf \it light} mass region A$_{CN}$ $\leq$ 60, the
fusion-fission (FF) process plays an important role in the compound nucleus
(CN) decay \cite{Sa86,Be92,Sa87,Ra91,Sa91,Ma91,Dj92,Be93,Be96,Fa96}. One of the
difficulties in this light-mass region is that the fully-damped yields of most
of the observed binary-decay products are mixed with those of quasi-elastic as
well as deep-inelastic processes and therefore their distinction from FF yields
is a rather difficult task for the experimentalist
\cite{Sa86,Sh87,Be92,Sa87,Be93,Be96,Fa96}.

\medskip

For the lighter mass systems, the nuclear orbiting process induced by a
long-lived dinuclear molecular complex, which subsequently binary decays, is
among the possible mechanisms of producing complex-fragments for which the
energy degree of freedom has been fully relaxed \cite{Sh87}. However, the
experimental data for the $^{16}$O+$^{40}$Ca \cite{Sa86}, $^{32}$S+$^{24}$Mg
\cite{Sa87}, $^{35}$Cl+$^{12}$C \cite{Be92}, $^{31}$P+$^{16}$O \cite{Ra91} and
$^{23}$Na+$^{24}$Mg \cite{Be93} reactions have been found to be consistent with
an equilibrated CN formation which subsequently binary-decays with the emission
of complex fragments i.e. a FF process. The occurrence of FF rather than
orbiting in these systems has been the subject of much discussion. This has led
to the conclusion that the FF process has to be taken into account when
exploring the limitations of the complete fusion process at large angular
momenta and high excitation energies \cite{Sa91,Be93}. Fig.1 and Fig.2
illustrate for the $^{32}$S+$^{24}$Mg \cite{Sa87} and $^{35,37}$Cl+$^{12}$C
\cite{Be92} reactions respectively two typical examples of sets of data which
have been selected to be compared with the results of the statistical model
developed in the present paper.

\medskip

The Extended Hauser-Feshbach Method (EHFM), which has been already presented in
several communications \cite{Ma91,Ma84}, assumes that the fission probability
is taken to be proportional to the available phase space at the scission point.
The EHFM corresponds essentially to an extension of the Hauser-Feshbach
formalism \cite{Ha52} which treats gamma-ray decay, light-particle evaporation
and complex-fragment emission (or FF) as the possible decay channels in an
equivalent way. In this paper we will apply the EHFM to {\bf light} heavy-ion
fusion-fission reactions. This is an alternative approach to the
transition-state model \cite{Sa91} using the phase space at the saddle point
which has provided quite good predictions of the available experimental data
\cite{Sa86,Sa87,Sa91,Dj92,Be93,Be96,Fa96}. Since there are good indications for
the validity of the hypothesis that the saddle-point shape almost coincides
with the scission-point configurations in the light-mass region, it is expected
that the EHFM might also be relevant. Preliminary results of EHFM calculations
as performed for the $^{35}$Cl+$^{12}$C fission reaction in Ref.\cite{Be96} are
quite conclusive.

\medskip

This paper is organized as follows : in the next Section, the essential points
of the EHFM are first presented. After a brief description of the well-known
Hauser-Feshbach formalism which is used for the CN light-particle emission, the
main characteristics of the formal procedures of the EHFM are described in
Section III. The complete EHFM+CASCADE calculation is applied for a
first-chance fission (or emission of excited complex fragments) followed by
their light-particle sequential decays until the resulting products are unable
to undergo further decay. This can be considered as a reasonable assumption for
light-mass systems as shown previously for a medium light-mass fission reaction
\cite{Na92}. In Section IV, the calculated results are shown in the case of a
simple parametrization and their applicability to the two selected reactions
studied \cite{Be92,Sa87,Be96} is discussed (cross sections are plotted in
Figs.1 and 2). Results are summarized and some preliminary conclusions are
finally drawn in Section V along with a short discussion relative to future
directions for systematic investigations and applications of the model to a
wider mass-range of nuclear systems from the light-mass region to the
intermediate-mass region.

\bigskip

\centerline{\bf II. EXTENDED HAUSER-FESHBACH METHOD}

\bigskip

In order to clarify the essential viewpoints of the EFHM calculations, the
salient formulae used in the well-known statistical model treatments which are
based upon the Hauser-Feshbach formalism \cite{Ha52} to describe the CN
light-particle evaporation are presented in next subsection . Given that the
statistical model \cite{St84}, which follows from the assumption of
equilibrium, rests on the premise that all open decay channels are, on the
average, equally likely to be populated it was natural to extend its formalism.
Although the treatment of light-particle emission and FF are, in principle,
inconsistent for heavier nuclear systems, it has been shown that in the case of
lighter nuclei (where the fissility parameters are below the Businaro-Gallone
point \cite{So84}) the asymmetric fission process can be assimilated to the
emission of larger fragments \cite{Mo75} which are also known as complex
fragments (or IMF's). Light-particle evaporation and FF, which are the two
commonly observed CN decay modes, appear to be just two extremes of a more
general binary-decay mode involving the entire range of mass-asymmetry
\cite{Mo75}. The extension of the Hauser-Feshbach method \cite{Ha52} to the
complex-fragment emission and/or FF is explained in subsection II.2 within the
framework of all available phase space. A brief description of the
parametrization of the transmission coefficients and its approximations are
given in subsection II.3. The complete calculation procedures of EHFM+CASCADE
which take into account the sequential emission of light-particles and
gamma-rays from the excited fission (or complex) fragments are finally
presented in Section III.

\bigskip

\centerline{\bf II.1 - Hauser-Feshbach Method for light-particle evaporation.}

\medskip

Most of the commonly used statistical-model codes, such as CASCADE \cite{Pu77},
PACE \cite{Ga80} or LILITA \cite{Go81} which have provided good predictive
results for the evaporation residues (ER) yields measured for a large number of
fusion-evaporation reactions, are based on a method proposed by Hauser and
Feshbach more than four decades ago \cite{Ha52}.

In the Hauser-Feshbach method the cross section ${\sigma^{(c)}_J}$ for the CN
formation and its subsequent statistical decay to channel ${c}$ whose state is
populated at an excitation energy E$_{x}$ with a total angular momentum ${J}$
is given by using the decay ratio ${R^{(c)}_J}$ as follows,

\begin{equation}
   {\sigma^{(c)}_J} = {R^{(c)}_J}{\sigma_J(E_x)} \hspace{2mm},
\label{eqn:f1}
\end{equation}
where ${\sigma_J(E_x)}$ is the cross section of the populated compound
states. Generally the ratio ${R^{(c)}_J}$ is determined by the ratio of
the partial width ${\Gamma_J^{(c)}}$ to the total width ${\Gamma_J}$ ,
\begin{equation}
   {R^{(c)}_J} = {{\Gamma_J^{(c)}} \over {\Gamma_J}} \hspace{2mm},
\label{eqn:f2}
\end{equation}
where the total width is a sum of all the partial widths of the decaying
channels ${c}$,
\begin{equation}
   {\Gamma_J} = \sum_c {\Gamma_J^{(c)}} \hspace{2mm} .
\label{eqn:f3}
\end{equation}
In the case of light-particle evaporation, the decay channel ${c}$ includes
mainly neutron, proton, $\alpha$-particle channels and $\gamma$-ray emission.
In some cases it can be also interesting to include $^{2,3}$H, $^{3}$He and Li
emissions \cite{Pu77}, but their influence is found to be negligible for the
reactions studied. In this paper these $^{2,3}$H and $^{3}$He channels are
only included in the calculations in the first-chance decay. Whereas the Li
channel is included and is considered as a complex-fragment emitted by a binary
decay in the phase-space calculation of EHFM as it will be explained more in
detail in the next subsection.

The partial width ${\Gamma_J^{(c)}}$ is related to a phase-space integration
${P_J^{(c)}}$,
\begin{equation}
   {\rho_J(E_x)} {\Gamma_J^{(c)}} = {1\over {2\pi}}{P_J^{(c)}} \hspace{2mm},
\label{eqn:f4}
\end{equation}
where ${\rho_J(E_x)}$ is the level density of the compound state. This level
density is not so relevant for the calculation of the decay ratio
${{R^{(c)}_J}}$ in eq.(\ref{eqn:f2}), but has a real physical meaning for the
estimation of a mean lifetime ${\tau_J}$ of the compound states. The mean
lifetime ${\tau_J}$ of a compound nucleus is generally evaluated by using the
total width ${\Gamma_J}$ as follows,
\begin{equation}
   {\tau_J} = {\hbar \over {\Gamma_J}} \hspace{2mm}.
\label{eqn:f5}
\end{equation}
This definition will be of interest in the discussion on the differences
between the lifetime of the compound states with the time needed to emit a
complex fragment near the scission point.

In the well-known Hauser-Feshbach method \cite{Ha52} used to describe the
light-particle emission, the phase-space integration ${P_J^{(c)}}$ to the
channel c is evaluated by the following phase-space integration,

\begin{equation}
   {P_J^{(c)}} = g_c \sum_{(L,I)J} \int \int  \rho_I (\epsilon)T_L (E)\delta
 (\epsilon+E+Q-E_x) d\epsilon dE \hspace{2mm}.
\label{eqn:f6}
\end{equation}

Here ${g_c}$ denotes the spin multiplicity of the evaporated particle and
${\rho_I(\epsilon)}$ is the level density of the residual nucleus with internal
excitation energy ${\epsilon}$ and angular momentum ${I}$. ${T_L(E)}$ are the
transmission coefficients for the evaporated particles as a function of energy
${E}$ and angular momentum ${L}$ in the relative motion with the daughter
nucleus. We use the transmission coefficients obtained in the optical-model
(OM) calculations in which the potential parameters have smooth dependences on
the mass number and are standard in the statistical-model calculations
\cite{St84,Pu77}. The ${(L,I)J}$ shown in the summation of angular momentum
represents the proper angular momentum coupling condition. The energy
conservation condition is maintained by ${\delta (\epsilon+E+Q-E_x)}$ in
eq.(\ref{eqn:f6}).

As soon as the excited states of the daughter nucleus are low enough in energy,
which is normally the case at the end of the cascade calculations, it is
necessary to take into account in the phase-space integrations of the
light-particle decay the experimentally known discrete levels of the daughter
nucleus near its ground state. As a consequence the phase-space integration
${P_J^{(c)}}$ becomes a summation of the known discrete levels ${i}$ of the
daughter nucleus, instead of the energy integration with ${\epsilon}$ in
eq.(\ref{eqn:f6}),
\begin{equation}
   {P_J^{(c)}} = g_c \sum_{i} \sum_{(L,I_i)J} \int T_L (E)\delta
 (\epsilon_i+E+Q-E_x)  dE \hspace{2mm},
\label{eqn:f7}
\end{equation}
where ${\epsilon_i}$ and spin ${I_i}$ are the known ${i}$-th discrete levels of
the daughter nucleus which have been taken from recent compilations
\cite{Aj91,En90}. For the emission of ${\gamma}$-rays, we include only the
giant dipole resonance (GDR) decay by using the form factor of Ref.\cite
{Cst87}.

The quantity Q is the usual separation energy for the light-particle
evaporation which is defined as follows,
\begin{equation}
  Q = B_{GS}(N_{CN},Z_{CN}) - B_{GS}(N_{L},Z_{L}) - B_{GS}(N_{H},Z_{H})
\hspace{2mm},
\label{eqn:f8}
\end{equation}
where ${B_{GS}(N_{CN},Z_{CN})}$,${B_{GS}(N_{L},Z_{L})}$ and
${B_{GS}(N_{H},Z_{H})}$ are the binding energies of the CN, evaporated particle
and daughter nucleus, respectively. The observed ground-state binding energies
given by the data tables \cite{Wa85} for the evaluation of Q-values are used.
If we include for example $^{3}$He evaporation in the calculations, it is found
to be negligible in agreement with experiment results since $^{3}$He has a very
small binding energy if compared to the $\alpha$-particle binding energy.

For the evaluation of the level density the Bohr and Mottelson expression
\cite{Bm69}, which is derived from the Fermi-gas model, has been used,
\begin{equation}
\rho_I (\epsilon)={1\over {12}}\Bigl( {{a\hbar^2}\over{2{\cal J}}}\Bigr) ^{3/2}
(2I+1)a{{e^{2\sqrt{X}}}\over{X^2}}  \hspace{2mm}.
\label{eqn:f9}
\end{equation}
where
\begin{equation}
X = a(\epsilon - {{\hbar^2}\over{2{\cal J}}}I(I+1)-\Delta_{pair} )
\hspace{2mm}.
\label{eqn:f10}
\end{equation}
The ${\cal J}$ is the moment of inertia of the daughter nucleus. In this
paper we use the well-parametrized moment of inertia of spherical nucleus
shown in the Ref. \cite{Bm69},
\begin{equation}
{\cal J} = {2 \over 3} A M {<r>_A^2} \hspace{2mm};
 \hspace{1cm} {<r>_A^2} = {3 \over 5}(1.12A^{1/3})^2(1+3.84A^{-1/3})
 \hspace{2mm},
\label{eqn:f11}
\end{equation}
where ${M}$ is the nucleon mass and ${<r>_A^2}$ is the mean square radius of
the ground state of the nucleus. For the sake of simplicity, we use a constant
level density parameter value. For the calculations we have chosen the value
${a=A/8}$ which appears to be rather well established both experimentally
\cite{St84,Pu77,Fo91} and theoretically \cite{Sh91} for the light heavy-ion
systems considered in the present study. The pairing energy ${\Delta_{pair}}$
is given by the empirical value ${\Delta_{pair} = 12/\sqrt{A}}$ as proposed in
Ref.\cite{Bm69}.

EHFM calculations have been performed for previously studied complete-fusion
reactions in the A$_{CN}$ $\approx$ 30 \cite{Be85} and A$_{CN}$ = 56
\cite{Be89a} mass regions in order to test the predicting capabilities of the
present model for the fusion-evaporation residues. Their comparisons with the
data \cite{Be85,Be89a} and with predictions of the evaporation codes CASCADE
\cite{Pu77}, PACE \cite{Ga80} or LILITA \cite{Go81} clearly show that first the
number of evaporated light-particles is correctly predicted by the EHFM and,
secondly that the results are not too sensitive to the choices of the
approximations and of the parameters of the present model.

\bigskip

\centerline{\bf II.2 - Extension of the Hauser-Feshbach formalism to the
binary-decay.}

\medskip

The objective of the EHFM is to extend the Hauser-Feshbach formalism
\cite{Ha52}, which has been described previously, to the phase-space
integrations of the binary-decays of the complex fragments (or the fission
decay width) from the compound nucleus. The phase-space integrations for the
complex-fragment binary-decays consist of four parts which are defined by the
four forthcoming equations.

\smallskip

At first we consider the case of the binary decays in which the lighter partner
of binary pair is populated in the discrete levels at low energies near the
ground state and the heavier one is in higher excitation-energy states in the
continuum region. The phase-space integration ${P_J^{(c)}}$ for this binary
decay is then assumed to be evaluated by the extension of the eq.(\ref{eqn:f6})
as follows,
\begin{equation}
   {P_J^{(c)}} = \sum_i \sum_{(I_{L_i},I_H)I}~~ \sum_{(L,I)J} \int \int
  \rho_{I_H} (\epsilon_H)T_L (E)\delta (\epsilon_{L_i}+\epsilon_H+E+Q-E_x)
 d\epsilon_H dE \hspace{2mm},
\label{eqn:f12}
\end{equation}
where ${\rho_{I_H} (\epsilon_H)}$ is the Fermi-gas level density of the heavier
fragment with excitation energy ${\epsilon_H}$ and angular momentum ${I_H}$.
${\epsilon_{L_i}}$ and ${I_{L_i}}$ are the excitation energy and angular
momentum of the i-th discrete levels of the emitted light fragments. In the
present calculation we introduce the known low excited discrete levels
\cite{Aj91,En90} in the low-energy region of each binary-decay fragment of
interest up to the lowest particle decay threshold
energy. The calculated results performed without including discrete levels
badly reproduce both the yields and the energy distributions which are known to
exhibit a structure understood in terms of the statistical population of levels
in the fragments \cite{Fa96}.

As in the case of light-particle emission, in this paper the level density is
calculated by using the moment of inertia spherical nucleus shown in
eq.(\ref{eqn:f11}), thus the deformation effects are not introduced in the
level density of both lighter and heavier fragments in the binary decay. This
possibility will be investigated in a subsequent publication \cite{Be97}by
including the angular-momentum-dependent terms in the ground-state moment of
inertia as proposed recently by Huizenga et al. \cite{Hu89}.

In a similar manner to light-particle evaporation, E is the energy of the
relative motion between the lighter fragment and heavier binary partner,
${T_L(E)}$ is the transmission coefficient of the relative motion with a given
angular momentum ${L}$. As we are trying to extend the framework of the
Hauser-Feshbach method of the light-particle emission to the case of complex
fragments emission, it is more reasonable to introduce the transmission
coefficients obtained in the OM calculation for evaluating the transmission
coefficients. However, in this study we will use a simplified formula for the
transmission coefficient as will be explained in the following subsection. As
will be discussed in the Section IV, the phase-space integration of
eq.(\ref{eqn:f12}) will mainly contribute to the mass-asymmetric binary-decay.
Phase-space calculations of this kind have already been extended to the study
of the emission of complex fragments in the case of the $^{58}$Ni+$^{58}$Ni
reaction with quite reasonable success in predicting complex-fragment charge
distributions that have been experimentally measured by the Oak Ridge group
\cite{Go91}.

Next we apply the above considerations to the case of the light fragments
highly excited in the continuum-energy region. Instead of the summation up to
the i-th discrete levels, the integration in the excitation energy
${\epsilon_L}$ and summation of angular momentum ${I_L}$ of the light fragment
is performed as follows,

\begin{equation}
   {P_J^{(c)}} =  \sum_{(I_L,I_H)I}~~ \sum_{(L,I)J} \int \int \int
 \rho_{I_L}(\epsilon_L) \rho_{I_H}(\epsilon_H) T_L (E)\delta
 (\epsilon_L+\epsilon_H+E+Q-E_x) d\epsilon_L d\epsilon_H dE \hspace{2mm},
\label{eqn:f13}
\end{equation}
where ${\rho_{I_L}(\epsilon_L)}$ is the Fermi-gas level density of the light
fragment. In order to integrate this large phase space a long computational
time is necessary. This is however the most essential part of the EHFM which
are applied to the mass-symmetric part of FF.

The phase-space calculation for the heavier fragment in the low excitation
energy region with discrete states in a similar manner to
eq.(\ref{eqn:f12}) is as follows,
\begin{equation}
   {P_J^{(c)}} = \sum_j \sum_{(I_L,I_{H_j})I}~~ \sum_{(L,I)J} \int \int
 \rho_{I_L} (\epsilon_L)T_L (E)\delta (\epsilon_L+\epsilon_{H_j}+E+Q-E_x)
 d\epsilon_{L_j} dE \hspace{2mm},
\label{eqn:f14}
\end{equation}
where the ${\epsilon_{H_j}}$ and ${I_{H_j}}$ denote the excitation energy and
angular momentum of the j-th discrete level of the heavier partner of binary
decay.

In the case where the fragments are both excited in the low-energy region, the
phase space is evaluated following eq.(\ref{eqn:f7}) which corresponds to the
phase-space integration of light-particle evaporation,
\begin{equation}
   {P_J^{(c)}} = \sum_i \sum_j \sum_{(I_{L_i},I_{H_j})I}~~ \sum_{(L,I)J} \int
  T_L (E)\delta (\epsilon_{L_i}+\epsilon_{H_j}+E+Q-E_x)  dE \hspace{2mm} .
\label{eqn:f15}
\end{equation}

As shown above, the phase-space integration for the complex-fragment
binary-decays consists of the four parts which are represented in
eqs.(\ref{eqn:f12}),(\ref{eqn:f13}),(\ref{eqn:f14}) and (\ref{eqn:f15}). In the
actual calculation, in order to avoid any possible overcounting, the
continuum-energy integrations for the level density formulae in
eqs.(12),(13),(14) are performed in the energy region starting from the energy
which is higher that the hightest excitation energy of the discrete levels
which are introduced in the discrete level summations of eq.(15). As mentioned
previously, the available discrete levels are taken below the lowest separation
energy in the neutron, proton and alpha-particle separations of the fragments.
For the evaluation of Q-values in these calculations, the observed ground-state
binding energies \cite{Wa85} are correctly used as shown in eq.(\ref{eqn:f8})
to keep the energy conservation condition. This effect which is clearly visible
in Fig.1.(c) will be discussed in the Section IV.

\bigskip

\centerline{\bf II.3 - Parametrization of the transmission coefficients. }

\medskip

As mentioned in the previous subsection, it would be highly desirable to use
explicitly the transmission coefficients of the OM calculations as a natural
extension of the Hauser-Feshbach method to the case of the complex-fragment
emission. However, because of the limitations of the computational time needed
to perform OM calculations, the transmission coefficients for these phase-space
integrations in eqs.(\ref{eqn:f12}),(\ref{eqn:f13}),(\ref{eqn:f14}) and
(\ref{eqn:f15}) are evaluated by using the simplified formula,
\begin{equation}
T_L (E) ={{1}\over{1+\exp ((V(L)-E)/\Delta_s)}} \hspace{2mm},
\label{eqn:f16}
\end{equation}
where the parameter ${\Delta_s}$ is the diffuseness parameter in the
transmission coefficient formula whose value has been kept equal to 0.5 MeV in
this study. This choice is consistent with the larger 1 MeV value which has
been recently chosen for the EHFM description of a heavier mass system
\cite{Na92}. As far as we know, the energy dependence of the transmission
coefficients obtained by the OM calculation are roughly fitted by the formula
(\ref{eqn:f16}) and the chosen diffuseness parameter value is comparable to
that of OM calculations in light-mass systems in the low angular momentum
region. For evaluating the transmission coefficients by the OM, the real part
has been deduced from fits to the measured elastic scattering cross sections if
available. In the case of the $^{35}$Cl+$^{12}$C scattering the OM parameter
set extracted from the elastic data measured by Djerroud \cite{Dj92} has been
used. The imaginary part is modified by the inclusion of short-range and sharp
diffuseness in order to reproduce the energy dependence of measured fusion
cross sections for light-mass systems in the so-called first regime of fusion
just above the Coulomb barrier. Of course the diffuseness for larger angular
momenta in OM calculations becomes much larger with increasing angular
momentum. We use the diffuseness parameter which is independent with the
angular momentum in this study. The calculations using the transmission
coefficients of the OM calculations will be discussed in a forthcoming paper
\cite{Be97}.

In order to simplify the discussions of the calculated results which will be
given in this paper, we have adopted the simple parametrization of the barrier
height ${V(L)}$ at the scission point between complex fragments which has been
assumed in the case of the $^{35}$Cl+$^{12}$C FF reaction {\cite{Dj92}} to be,
\begin{equation}
V(L) = V_{coul} + {{\hbar^2}\over {2\mu_f R^2_s}}L(L+1) \hspace{2mm},
\label{eqn:f17}
\end{equation}
where ${\mu_f}$ is the reduced mass of the decaying complex fragments. The
scission point ${R_s}$ is estimated by using the radius ${R_L = r_s A_L^{1/3}}$
and ${R_H = r_s A_H^{1/3}}$ of the two fragments of mass number ${A_L}$ and
${A_H}$ including diffuse-surface effects with a neck length parameter (or
separation distance) ${d}$,
\begin{equation}
R_s = R_L + R_H + d \hspace{2mm},
\label{eqn:f18}
\end{equation}
and the ${V_{coul}}$ is calculated by the following simple formula,
\begin{equation}
V_{coul} = {Z_LZ_H e^2}/{R_s} \hspace{2mm},
\label{eqn:f19}
\end{equation}
where ${Z_L}$ and ${Z_H}$ are the atomic numbers of the lighter and heavier
exit-fragments respectively. The neck length parameter ${d}$ is taken as the
only adjustable parameter of the model. Its value is found to be ${d}$ = 3.0
$\pm$ 0.5 fm as is commonly adopted in the literature
\cite{Be92,Dj92,Na77,Va83,Pu96} for the mass region of interest. The large
value of ${d}$ used for the neck mimics the finite-range and diffuse-surface
effects \cite{Sa91} of importance for the light-mass systems \cite{Bes96} and,
as a consequence, this makes the scission configurations closely resemble the
saddle configurations. The other parameters are either fixed (for instance we
use a constant value of ${r_s=1.2 fm}$ in this work in accordance with previous
studies \cite{St84,Pu77}) or determined by the measured fusion cross sections
(see Section III). An alternative and more sophisticated approach to evaluate
the transmission coefficients at the scission point is the use of Krappe, Nix
and Sierk \cite{Kr79} potential for V(L=0). Calculations of this kind have been
performed for the $^{35}$Cl+$^{12}$C reaction \cite{Be92,Dj92} at E$_{lab}$ =
180 and 200 MeV with very similar results as the ones shown in Fig.2 with the
simplest parametrization. Another study \cite{Au87} involving a heavier mass
system has shown that the choice of the potential does not provide very
different predictions in statistical models.

All other quantities for evaluating the phase-space integrations in
eqs.(\ref{eqn:f12}),(\ref{eqn:f13}),(\ref{eqn:f14}) and (\ref{eqn:f15})
are the same as in the case of the light-particle evaporation description.
In the actual calculations, the phase-space integrations of
eqs.(\ref{eqn:f12}),(\ref{eqn:f13}),(\ref{eqn:f14}) and (\ref{eqn:f15}) are
performed with a high precision for the  numerical integrations without any
approximations. The energy integrations are performed with 1 MeV energy steps
and all the values of the cascade decay are stored in 1 MeV steps and 1
${\hbar}$ angular momentum steps.

\bigskip

\centerline{\bf III. EHFM+CASCADE CALCULATION PROCEDURES }

\bigskip

In order to show the basic viewpoint of the EHFM and to demonstrate its
applicability for the decay mechanisms of compound nuclei as formed in light
heavy-ion reactions, we perform the EHFM+CASCADE calculations by introducing
the very simplified schemes which are presented as follows. In light heavy-ion
reactions at moderate incident energies such as the $^{32}$S+$^{24}$Mg and
$^{35}$Cl+$^{12}$C reactions which have been selected as typical examples in
this paper, it seems quite reasonable to assume that the second-chance, the
third-chance and many-chance binary-decays from the heated daughter nuclei
(ER's) populated by light-particle evaporation in the early stages are
negligible. Therefore the EHFM is applied to the phase-space calculations only
in the first-chance binary-decay of complex-fragments (i.e., first-chance
fission-like binary-decay) from the fused system. Because the complex fragments
emitted in the first-chance decay are expected to be populated in the rather
highly excited states in both energy and angular momentum in a similar manner
to the ER's, the heated fragments including the ER's need to be cooled down
until the resulting products are unable to undergo further decay. Therefore the
series of these calculations is called as EHFM+CASCADE.

From a result of the EHFM+CASCADE calculations physical quantities such as
charge-, mass- and kinetic-energy distributions can be deduced to be compared
to experimental observables. For instance the missing charge and its
distributions corresponding to the experimental conditions in coincidence
measurements are found to be clearly described with this EHFM+CASCADE
calculation. These definitions are found to reasonably well reproduce the
experimental distributions as shown in the next Section. In the present
Section, the calculation procedures of EHFM+CASCADE are presented according to
the chosen approximations.

The initial conditions required to perform the EHFM+CASCADE calculation are
mainly determined by the total fusion cross section ${\sigma_{fus}}$  which is
assumed to be given as the compound nucleus formation with atomic number
${Z_{CN}}$ and neutron number ${N_{CN}}$ and at the excitation energy ${E_x}$
in the heavy-ion reaction under consideration as follows,
\begin{equation}
 {\sigma_{fus}} = \sum_{J=0}^{\infty} {\sigma_{fus}(J)}
 = {\pi}{\lambar}^2 \sum_{J=0}^{\infty} (2J+1){T^{(fus)}_{(J)}} ,
\label{eqn:p1}
\end{equation}
where ${\lambar}$ and ${J}$ are respectively the wave length and the total
angular momentum of the incident channel of the reaction. For the sake of
simplicity, the partial wave dependence of the fusion cross section
${\sigma_{fus}(J)}$ are represented by the transmission coefficient
${T^{(fus)}_{(J)}}$ with Fermi distribution,
\begin{equation}
T^{(fus)}_{(J)}={{1}\over{1+\exp((J-{J_{cr}})/\Delta_J)}} .
\label{eqn:p2}
\end{equation}
The critical angular momentum ${J_{cr}}$ is chosen so as to reproduce the
measured complete fusion cross section ${\sigma_{fus}}$ including both ER and
FF yields. Although little is known about the diffuseness parameter
${\Delta_J}$, its value has been fixed to ${1 \hbar}$ in the present study in
accordance with the value usually taken for the transition-state model
calculations of Sanders \cite{Sa91} or other evaporation codes
\cite{Pu77,Ga80}. The sensitivity of this angular-momentum diffuseness
parameter has been carefully checked and very small and thus non significant
effects have been found on the calculated results.

\medskip

All the available complex-fragment pairs are introduced in the first-chance
decay as part of the binary-decay in addition to proton, neutron and
$\alpha$-particle evaporation (ER's) from the fused system. As discussed in the
previous Section, the $^{2,3}$H and $^{3}$He evaporation channels are
included only in the first-chance decay, whereas the GDR ${\gamma}$-ray
emission is also included in the whole CASCADE-calculation.

The decay ratio ${R_J^{(c)}}$ in eq.(\ref{eqn:f2}) is evaluated, as shown
previously, for all of the exit-channels by using the fusion partial cross
section ${\sigma_{fus}(J)}$ of eq.(\ref{eqn:p1}) as the cross section
${\sigma_J{(E_x)}}$ of eq.(\ref{eqn:f1}). During the course of the calculations
of the whole phase-space integrations for all decay channels, all the
quantities which are needed in the subsequent CASCADE-calculations should be sto
However, due to the fact that the memory space of the available computer is not
large enough to store the calculated results for all the dependent variables in
the first-chance EHFM calculation, the calculated results are stored in the two
groups somewhat inclusively as follows. The excitation energy and angular
momentum distributions for each fragment with atomic number ${Z'}$ and neutron
number ${N'}$ are stored in the form of ${\sigma_{(Z',N')}(\epsilon',I')}$ as a
value of cross section and the kinetic-energy distribution of the first-chance
emission of fragments is also stored as the form ${\sigma_{(Z',N')}(E')}$ where
the E is the kinetic energy of the relative motion between the
binary-fragments. In order to make the notations clear, in this paper the
superscript of prime is put on the quantities which correspond to the
first-chance EHFM calculations. As has been expected, the distributions
${\sigma_{(Z',N')}(\epsilon',I')}$ of the fragments with atomic number ${Z'}$
and neutron number ${N'}$ obtained in the first-chance decay are in rather
highly excited states (for example, see Fig.3 as discussed in the next
Section). Then the usual CASCADE-calculations are applied to the hot fragments
thus populated in the first-chance EHFM calculation.

\medskip

The final distributions ${\sigma_{(Z',N')}(\epsilon,I,Z,N)}$ of the fragment
with atomic number ${Z}$ and neutron number ${N}$ are obtained as the result of
the light-particle cascade-decay of each fragment with atomic number ${Z'}$ and
neutron number ${N'}$ which is populated with the cross section
${\sigma_{(Z',N')}(\epsilon',I')}$ in the first-chance EHFM calculation. In the
course of the CASCADE-calculation, the distributions
${\sigma_{(Z',N')}(\epsilon,I,Z,N)}$ can be stored in the computer memory, but
due to computer memory limitations the final results are stored in the
inclusive form :
\begin{equation}
 \sigma_{(Z',N')}(Z,N) =  \sum_{\epsilon,I}
                        {\sigma_{(Z',N')}(\epsilon,I,Z,N)}
\label{eqn:p21}
\end{equation}
Then the charge- and mass-distributions, ${\sigma(Z)}$ and ${\sigma(A)}$ can be
directly compared to experimental data by summing up the final distributions
${\sigma_{(Z',N')}(Z,N)}$ relative to each the first-chance emitted fragment
with atomic number ${Z'}$ and neutron number ${N'}$.

\medskip

The average velocities of the first-chance emitted fragments are not expected
to be greatly modified by the effect of post-scission light-particle
cascade-decay. The fragment kinetic-energy distributions
${\sigma_{(Z',N')}(E')}$ which are obtained in the first-chance EHFM
calculation in the center-of-mass system are transformed to kinetic-energy
distributions for a given laboratory angle ${\theta_{lab}}$
\begin{equation}
   {d^2{\sigma_{(Z',N')}} \over {{d \Omega_{lab}} {d E'_{lab}}}}
\label{eqn:p3}
\end{equation}
by using the usual transformation formula. In this calculation the angular
distribution of the fragments of binary decay at the first-chance emission is
assumed to have the usual ${1/{sin(\theta_{cm})}}$ angle dependence in the
center-of-mass system of fission-like processes.

In order to simplify the notation to define both the missing charge and the
kinetic energy distributions in the calculation, the charge distributions
${\sigma_{(Z',N')}(Z)}$ of the final fragments with the atomic number ${Z}$
which are populated by the light particle cascade-decay of the hot fragments
with atomic number Z and neutron number N is :
\begin{equation}
 \sigma_{(Z',N')}(Z) =  \sum_{N} {\sigma_{(Z',N')}(Z,N)}
\label{eqn:p4}
\end{equation}
and the probability distributions ${P_{(Z',N')}(Z)}$ are defined as a function
of the charge distributions ${\sigma_{(Z',N')}(Z)}$ as,
\begin{equation}
 P_{(Z',N')}(Z) =  {{\sigma_{(Z',N')}(Z)}\over {\sum_{Z"}\sigma_{(Z',N')}(Z")}}
\label{eqn:p5}
\end{equation}

By using the probability distributions ${P_{(Z',N')}(Z)}$, the kinetic-energy
distributions modified by the light particle cascade-decay after the scission
are evaluated as follows for the fragments with the atomic number ${Z'}$ and
the kinetic energy ${E'_{lab}}$ which have been obtained in the first-chance
EHFM calculation,
\begin{equation}
 {d^2{\sigma_{(Z)}} \over {{d \Omega_{lab}} {d E_{lab}}}}
 = \sum_{(Z',N')} P_{(Z',N')}(Z) {d^2{\sigma_{(Z',N')}} \over {{d \Omega_{lab}}
{d E'_{lab}}}}
\label{eqn:p6}
\end{equation}
Here the relation between kinetic energy ${E'_{lab}}$ of the first-chance
emission and final measured kinetic energy ${E_{lab}}$ is assumed to be used as
follows :
\begin{equation}
  E_{lab}= {Z \over Z'} E'_{lab} .
\label{eqn:p7}
\end{equation}

For the evaluation of the missing charge distribution corresponding to the
measured ejectile fragment with atomic number ${Z_1}$ in the coincidence
measurement (see Ref.\cite{Be96} for the experimental conditions and results),
the coincidence cross sections ${\sigma_{Z_1}^{(coin)}(Z_2)}$ are defined as
follows for the coincident binary partner with atomic number ${Z_2}$
\begin{equation}
 {\sigma_{Z_1}^{(coin)}(Z_2)} = \sum_{(Z'_1,N'_1)}
                               {\sigma_{(Z'_1,N'_1)}(Z_1)}
                               {\sigma_{(Z'_2,N'_2)}(Z_2)}
\label{eqn:p8}
\end{equation}
where we must keep the condition; ${Z'_2=Z_{CN}-Z'_1}$ and
${N'_2=N_{CN}-N'_1}$. Then the probability distribution of missing charge for a
first fragment with ${Z_1}$ is given in the following formula,
\begin{equation}
 P_{(Z_1)}(\Delta Z) =  {{\sigma_{(Z_1)}^{(coin)}(Z_2)}\over
 {\sum_{Z"}\sigma_{(Z_1)}^{(coin)}(Z")}}
\label{eqn:p9}
\end{equation}
where ${\Delta Z = Z_{CN} - (Z_1 + Z_2) }$ is defined as the missing charge.

Finally the mean values ${\langle Z_1+Z_2 \rangle}$ which correspond to the
measured mean charge in coincidence measurements are defined by using the
probability distributions ${P_{(Z_1)}(\Delta Z)}$ of missing charge as follows,
\begin{equation}
 {\langle Z_1+Z_2 \rangle} = \sum_{\Delta Z} (Z_1+Z_2) P_{(Z_1)}(\Delta Z)
\label{eqn:p10}
\end{equation}

In comparison with other recent statistical-model calculations
\cite{Go91,Sa91}, it is worthwhile to mention that one of the main advantages
of the present model is the use of a single computer code to follow the whole
decay process until all fragments have completely cooled down.

\newpage

\centerline{\bf IV. RESULTS AND DISCUSSIONS}

\medskip

Before we present the results of the EHFM+CASCADE calculations which have been
performed for a few selected examples, it is important to notice that in this
light-mass region it is relevant to use the scission-point approximation of the
saddle point. This is due to the fact that both the scission point and the
saddle point have geometrical configurations which nearly coincide as recently
demonstrated \cite{Be96} in the case of the binary-decay of the $^{47}$V
system. Alternative available computer codes such as EDCATH \cite{Au87}, GEMINI
\cite{Ch88} or EUGENE \cite{Du92} are essentially based on the saddle-point
picture by using the transition-state formalism of Moretto \cite{Mo75} to
predict complex-fragment emission yields for heavier systems. The
transition-state model developed by S. Sanders \cite{Sa91} and more
specifically adapted for the light-mass region appears to be quite successful
by introducing mass-asymmetric fission barriers. On the other hand the code
BUSCO \cite{Go91} is to our knowledge the only code also following the
scission-point approximation with, however \cite{Go96}, the need of the code
LILITA \cite{Go81} to simulate the sequential decay of the binary fragments.

In this Section the results of the model will be compared to a number of
recently published experimental data \cite{Be92,Sa91,Be96,Be89}. For a
more general overview of the experimental systematics of the occurrence of the
FF process in the light-mass region previous publications such as
Ref.\cite{Sa91} are very helpful.

\medskip

As pointed out previously in Section II one of the most important quantities in
the EHFM is the measured ground-state binding energy used to evaluate Q-values
for the all complex fragments in the phase-space calculations in order to
explicitly conserve energy. In order to demonstrate the strong
effect of the ground-state binding energy, first of all we choose as a typical
example the mass distributions as measured for the $^{32}$S+$^{24}$Mg
reaction at two incident energies E$_{lab}$ = 121 and 142MeV \cite{Sa87}
displayed in Fig.1.(a) and (b) respectively. It is very interesting to observe
that the calculated mass distributions shown by solid histograms reproduce well
the characteristic features of the variations from fragment to fragment in the
experimental mass distributions shown by open histograms. In these
calculations, the critical angular momenta for total fusion cross section at
the energies E$_{lab}$ = 121 and 142 MeV are respectively J$_{cr}$ = 34 and 37
$\hbar$. These values reproduce the measured complete fusion cross sections
which are reported in Ref. \cite{Sa87}. The free parameter ${d}$ in
eq.(\ref{eqn:f18}) which determines the barrier height of the scission point is
chosen in this case to be {\it d} = 3.5 fm. A systematic investigation of this
parameter will be undertaken in a forthcoming publication \cite{Be97}.

In order to understand the reasons why the calculated mass distribution is
strongly dependent on the ground-state Q-value of the decay fragments,
the fragment dependence of the barrier height of the scission point in
excitation energy of the compound nucleus for $^{56}$Ni is shown in
Fig.1.(c). As can be expected in the phase-space integrations ${P}$ which are
shown in eqs.(\ref{eqn:f12}),(\ref{eqn:f13}),(\ref{eqn:f14}) and
(\ref{eqn:f15}), the leading term can be evaluated approximately by the
following form,
\begin{equation}
 P \sim e^{2\sqrt{a(E_x - V_s)}}
\label{d1}
\end{equation}
where the value of {\it a} is equal to the sum of level density parameters of
the lighter and heavier fragments in the binary decay. The barrier height
${V_s}$ of the scission point which is evaluated from the ground state of the
CN is given as follows for the case of angular momentum ${L=0}$,
\begin{equation}
 {V_s} = V_{coul} + Q
\label{d2}
\end{equation}
In Fig.1.(c) the lowest barrier height of the scission point in the combination
of the same mass fragments with different atomic number are plotted. In the
case where the lighter fragment is a $\alpha$-like nucleus its heavy partner is
also a $\alpha$-like nucleus for the $^{56}$Ni system, therefore strong
binding energy effects are found in the barrier height of the scission point.
It is interesting to note that very similar results are found for the mass
fragmentation potential as calculated by Gupta et al. \cite{Gu84} for the
same system. On the other hand, it has been shown in Ref.\cite{Ma91} that with
the use of liquid-drop binding energies the yields do not vary significantly
from fragment to fragment. Comparing Fig.1.(a) and (b) with Fig.1.(c), the
strong enhancements in the measured cross section are well understood as the
result of the strong binding energy of $\alpha$-like fragments.

\smallskip

In the transition-state model calculation \cite{Sa87} the strong binding-energy
effect has been taken into account by including Wigner energy terms in the
liquid drop mass formula. Thus the origin of the strong variation from fragment
to fragment in the present model may be equivalent to that involved in the
transition-state model.

An alternative way to reproduce this strong variation would be to incorporate
shell effects in the level density formulae as proposed by Ignatyuk
\cite{Ig75}. Shell corrections in the energy-dependent (temperature-dependent)
{\it a} parameter are then produced by the difference of the experimental mass
and the liquid drop mass for each fragment. This possibility will be carefully
investigated in the future developments of EHFM \cite{Be97}; however
preliminary results on a study of the temperature-dependent level density can
be found in the conference proceedings \cite{Ma88}.

Despite of the choice of a very simple parametrization for the present
calculations, it should be pointed out that the complete EHFM+CASCADE treatment
reproduces well the general trend and also the magnitude of the measured mass
distribution. The calculated center-of-mass energy distributions which are
obtained in the first-chance emission of the EHFM calculation are found however
to be a little higher than the measured ones \cite{Sa87} by an amount of about
3 MeV.

\medskip

In the following we will focus on the case of the $^{35}$Cl+$^{12}$C reaction.
The calculated charge distributions in Fig.2 are compared to the experimental
data at E$_{lab}$ = 180, 200 and 278 MeV respectively as obtained in the
inclusive measurements \cite{Be92,Dj92,Be96,Be89}. The comparisons are also
given for the fission-like yields for the $^{37}$Cl+$^{12}$C which have been
partially measured at E$_{lab}$ = 150 MeV \cite{Yo89}. In these calculations
the input critical angular momenta J$_{cr}$ were extracted from the total
fusion cross section data using the sharp cutoff approximation. Their values
are 25 ${\hbar}$, 25 ${\hbar}$ and 27 ${\hbar}$ for the $^{35}$Cl incident
energies 180, 200 and 278 MeV respectively. Since the cross section of the
complete fusion ER has not been measured for the $^{37}$Cl+$^{12}$C system at
E$_{lab}$ = 150 MeV, a 23 ${\hbar}$ value was assumed to be the more realistic
choice. The value of the {\it d}-parameter for the barrier height of scission
point is fixed to be {\it d} = 2.5 fm for each incident energy. It is
interesting to observe that this {\it d} value is smaller than that in the case
of the $^{32}$S+$^{24}$Mg reaction. Although no systematics of the mass
dependence of the {\it d}-parameter for light mass-systems is evident for the
moment, it seems that a simple linear-dependence with the CN fissility might be
a reasonable assumption. This possibility will be further quantitatively
investigated within the framework of a more systematic study in a forthcoming
publication \cite{Be97}.

The charge distributions produced in the first-chance emissions by the EHFM
calculations are shown as dashed histograms in Fig.2 whereas the solid
histograms represent the final charge distributions obtained by performing
EHFM+CASCADE calculations. In the results of the EHFM calculations as the first
chance decay, we can clearly see the original traces of the binary pairs which
are introduced in these calculations as the available decay channels. The cross
sections with atomic number ${Z=}$ 20 and 19 arise from the Li and Be
emissions respectively within a binary-decay process. By comparing the cross
sections of complex-fragment binary-decays such as B and C emissions,
the Li and Be channels have significantly larger cross sections, but
these light complex-fragment binary-decays do not affect significantly the
largest part of the measured charge distributions which comprise the ER's. As
expected from the usual Hauser-Feshbach calculations, the cross sections with
charge Z = 20, 21 and 22 come mainly from the emissions of light particle such
as neutron, proton and alpha-particle. The $^{2,3}$H and $^{3}$He channels
included in the first-chance EHFM calculations are found also to have
relatively much smaller contributions in the cross sections for the ER's.
Therefore the EHFM+CASCADE calculations for each fragment including the ER's
after scission take only the neutron, proton and $\alpha$-particle emissions
into account.

Therefore it can be seen that the results from the EHFM+CASCADE calculations
reproduce well the whole measured charge distributions over the entire range of
mass-asymmetry from the low-mass region of complex-fragment emission (FF) to
the heavy-mass region of the evaporation residues (ER) for all incident
energies (see the solid histograms in Fig.2). Because the measured ground-state
binding energies are involved in these calculations also, the large yields with
${\alpha}$-like fragments are observed for the mass-asymmetric part of the
complex-fragment emission. But the heavier partner of the complex-fragment
binary decay are not $\alpha$-like in this system, then the cross sections with
mass symmetric fragments are not so significant as opposed to the case of
$^{32}$S+$^{24}$Mg reaction presented in previous paragraph.

In the region of the heavier fragments with atomic number larger than about 15,
the cross sections are mainly due to the ER's produced by light-particle
cascade-decays from the compound system. The comparison of the fragments
$12<Z<15$ was not possible because their data could not be extracted so
accurately from the experiment \cite{Be96}, due to a mixing with both quasi-
and deep-inelastic components. It is known however that the heavier partners of
the complex-fragment binary-decay have large cross sections ($\approx$ 10 mb)
which correspond to the cross sections of the lighter fragments but the yields
are obscured by the considerably larger ER cross sections ($\approx$ 100 mb for
each Z at E$_{lab}$ = 278 MeV for example). In the charge region ${Z<14}$, the
complex-fragment emissions become important and essential to reproduce the
experimental charge distributions. The light-particle cascade-decay of the
heavier fragments after scission increases with increasing incident energy.
Furthermore light-particle emissions from the lighter complex fragments after
scission are apparent in the case of the highest incident energy E$_{lab}$ =
278 MeV.

At this point it is important to notice that both the predictions of
EHFM+CASCADE and the transition-state model \cite{Sa91} provide a quite
satisfactory agreement of the general trends of the $^{35}$Cl+$^{12}$C
experimental excitation functions over the whole energy range explored as
demonstrated in Ref.\cite{Be96}. This might be a good indication of the
validity of the hypothesis that the scission point configurations, as assumed
in the present study, almost coincide with the saddle-point shape of the
transition-state picture \cite{Sa91}.

In order to show how the complex fragments are populated in the excited states
at the scission point, the internal excitation energy and angular momentum
distributions, ${\sigma(\epsilon')}$ and ${\sigma(I')}$, of $^{12}$C fragment
obtained in the first-chance EHFM calculation in the case of E$_{lab}$ = 278
MeV are shown in Fig.3.(a). The energy distribution ${\sigma(\epsilon')}$ is
obtained by summing up the angular momentum variable ${I'}$ of the fragment,
\begin{equation}
   {\sigma(\epsilon')} = \sum_{I'} {\sigma_{(Z',N')}(\epsilon',I')} ,
\end{equation}
and the angular momentum distribution ${\sigma(I')}$ is also obtained by
integrating the internal energy ${\epsilon'}$,
\begin{equation}
   {\sigma(I')} = \int {\sigma_{(Z',N')}(\epsilon',I')} {d \epsilon'}.
\end{equation}
where the ${\sigma_{(Z',N')}(\epsilon',I')}$ is the obtained cross section in
the first-chance EHFM calculation (see Section III). The distributions of the
partner nucleus $^{35}$Cl of the $^{12}$C fragment in the first-chance
emission are shown in Fig.3.(b). As can be understood from Fig.3.(a), the
$^{12}$C fragment is excited in the lower excited discrete levels, especially
the ground ${0^+}$ and first excited ${2^+}$ (4.44 MeV) states. The third large
peak corresponds to the ${3^-}$ (9.64 MeV) state whereas the smaller one
corresponds to the second ${0^+}$ state.

On the other hand, it is clearly seen that the fragment-partner $^{35}$Cl is
statistically excited to the continuous states (continuum) with smooth
distributions both in internal energy ${\epsilon'}$ and in angular momentum
${I'}$. This is a typical behavior of a complex-fragment statistical emission
from a equilibrated fused nucleus in the light-mass system region.

At the highest studied incident energy E$_{lab}$ = 278 MeV, the CN excitation
energy of $^{47}$V is ${E_x}$ = 84 MeV, the CN lifetime for light-particle
emission can be evaluated to correspond to roughly {6${\times}$10$^{-22}$} sec
by using the standard formula expressed in eq. \ref{eqn:f5}. As can be seen in
Fig.3.(b), on the other hand the averaged excitation energy of the partner
nucleus $^{35}$Cl of the $^{12}$C fragment is about 35 MeV. The lifetime
which corresponds to this excited $^{35}$Cl nucleus is about ten times longer
than that of the $^{47}$V compound state. Thus we can point out that the
proposed picture of light-particle cascade decay after scission of the excited
complex fragments obtained in first-chance EHFM calculation is relevant.

The calculated kinetic-energy distributions of each fragment
(5 $\leq$ Z $\leq$ 11) in the laboratory system which were evaluated with the
procedures outlined in Section III are shown in Fig.4.(a) and (b) along with
the data taken at 7$^{o}$ for two indicated incident energies of E$_{lab}$ =
278 and 180 MeV, respectively. The dashed lines are the kinetic-energy
distributions obtained in the first-chance EHFM calculations and the solid
lines are the kinetic-energy distributions including the effect of the
light-particle cascade decay after scission. As expected for a so-called
inverse kinematics reaction, the kinetic-energy distributions in the laboratory
system can be decomposed in two parts: a) a high-energy component with a
typical Gaussian shape which is well measured b) a lower-energy component which
is deformed by the experimental energy threshold. Therefore it should be taken
into consideration that the large deviations from the calculated distributions
with the data in the lower-energy parts come from the non-ideal experimental
conditions.

In the case of the highest measured incident energy E$_{lab}$ = 278 MeV
shown in Fig.4.(a), the effect of the secondary cascade decay of light
particles from the hot binary fragments is clear in contrast to the case of
the lower incident energy of E$_{lab}$ = 180 MeV shown in Fig.4.(b) for
which the differences between the two calculations is not significant.
Although for the two calculated kinetic-energy distributions the barrier
heights of the scission point which have been used appear slightly larger
than the measured ones, the calculations reproduce reasonably well the
general trend of the experimental data for both the mean values and the
associated widths.

In Fig.5, the missing charge distributions for the elemental fragments which
have been obtained in the coincident data \cite{Be96}, with the experimental
conditions of ${Z_1 \geq 5}$ and ${Z_2 \geq 5}$ with the optimum values of the
angular correlations, are shown by the solid histograms. Here the ${Z_1}$ is
the atomic number of the first fragment and ${Z_2}$ the atomic number of its
binary partner in the coincidence measurement. The distributions as calculated
with EHFM+CASCADE are shown by the dashed histograms for each fragment with
atomic number ${Z_1}$. The details of these calculations which were adapted to
the experimental conditions are given in Section III by assuming that the
missing charges have their origin from the cascade decay of light particles or
binary fragments after scission.

Despite the relative simplicity of the calculations resulting from the direct
use of the ground-state binding energy for the full course of the cascade decay
without any corrections for the level density parameters, the calculated
distributions for each coincident fragment reproduce well the general trend of
the experimental results. The large discrepancies observed for ${Z_1 = 5}$
might be due to an experimental bias arising from the geometry of detector
angles chosen for the coincidence measurements of the angular correlations. The
strong $\alpha$-emission in the calculated results are observed for the missing
charge distributions of the ${Z_1 = 11}$ and ${12}$ binary pair. These
deviations should be considered carefully in future studies taking into account
the ground-state binding energies directly in the calculations. However it
should be stressed that these calculated missing charge distributions reproduce
well the general trends of the experimental ones obtained in the coincidence
measurements \cite{Be96}.

Finally we can evaluate the average values of missing charges for the
distributions as shown in Fig.5, which results can be summarized in Fig.6 by
showing how the averaged values depend on the fragments of the coincidence
measurements by the use of the mean values ${\langle Z_1+Z_2 \rangle}$. The
calculated mean values obtained in EHFM+CASCADE are shown by the solid lines
whereas the experimental results are displayed by the data points with their
associated error bars for the $^{35}$Cl+$^{12}$C reaction at E$_{lab}$ =
200 \cite{Dj92} and 278 MeV \cite{Be96} as a function of the atomic number
${Z_1}$ of the first fragment in Fig.6.(a) and (b), respectively. As can be
seen in the comparisons between the experimental and calculated missing charge
distributions, it should be once again stressed that the EFHM+CASCADE
calculations reproduce well the general trend of the experimental data.

Knowing that the pre-scission emission of light particles are predicted to be
negligible in the model, it can be concluded that the missing charge obtained
for the $^{35}$Cl+$^{12}$C reaction in the coincidence measurements for
bombarding energies lower than 8 MeV/nucleon has its origin in the
light-particle cascade decay of the excited binary fragments after scission. A
similar conclusion has been advanced for study of the $^{35}$Cl+$^{24}$Mg
reaction measured at $\approx$ 8 MeV/nucleon which preliminary experimental
results are also well reproduced by EHFM+CASCADE \cite{No96}.

\bigskip

\centerline{\bf V. SUMMARY AND CONCLUSIONS}

\bigskip

In order to treat the binary-decay emission of complex-fragments in a similar
manner to the light-particle evaporation from light-mass compound systems as
populated by heavy-ion fusion reactions, the well-known Hauser-Feshbach
formalism has been extended in a natural way to the phase-space calculations of
the binary-decay (i.e. a fusion-fission process) (see Sec.II.2). The EHFM
calculation is applied to the first-chance binary-decay from the compound
system by assuming that the second-chance binary-decay (from the hot daughter
nuclei (ER's) populated by light-particle evaporation in the early stages) is
found to be negligible. The internal excitations of the emitted complex
fragments are populated in rather highly excited states in both angular momenta
and energies in the similar manner to the ER's. The hot binary fragments are
cooled down by the cascade of light-particle emissions. Subsequently
EHFM+CASCADE calculation can clearly define the physical quantities which are
able to be directly compared with the experimental ones such as missing charge
distribution (see Sec.III).

\medskip

The validity of these procedures is shown to be reasonable by referring to the
light-particle decay times which are found to have the expected values. In
order to make clear how the extension of the statistical model to the emission
of complex fragments is relevant, the calculations have been performed within
its most simplified version, namely the parameterization of the barrier height
of the scission point with the inclusion of the neck degree of freedom ${d}$
which mimics the diffuse-surface effects known to be of importance in the
light-mass region.

\medskip

The essential points of the EHFM+CASCADE calculation are presented with the
example of the binary-decay of the ${^{56}Ni}$ nucleus as formed by the
$^{32}$S+$^{24}$Mg reaction at two bombarding energies E$_{lab}$  = 121 and
142 MeV \cite{Sa87}. The EHFM+CASCADE calculation has also been applied to the
$^{47,49}$V systems which are formed in the $^{35,37}$Cl+$^{12}$C reactions at
E$_{lab}$ = 150, 180, 200 and 278 MeV \cite{Be92,Dj92,Be96,Be89,Yo89}. In this
case the neck length parameter of eq.(18) is fixed as {\it d} = 2.5 fm for all
incident energies, whereas its value has been found larger in the case of the
$^{56}$Ni compound system. As a matter of fact the neck degree of freedom
cannot be considered as a simple adjustable parameter since first its value d =
3.0 $\pm$ 0.5 fm appears to be strongly constrainted by the size, i.e. the
fissility of the compound system and secondly is non temperature-dependent for
a chosen reaction. Work is now in progress \cite{Be97} in order to define a
reasonable mass-dependence of this parameter through the investigation of new
available fusion-fission data on nuclei such as $^{44}$Ti \cite{Ol96},
$^{48}$Cr \cite{Fa96} or $^{59}$Cu \cite{No96}. The values of the critical
angular momenta ${J_{cr}}$ have been chosen so as to reproduce the measured
fusion cross sections. The post-scission light-particle decay of the emitted
complex fragments appears to be, in each studied case, of great importance to
obtain a reasonably good agreement for all the measured observables: mass-,
charge- and kinetic-energy distributions of both the complex fragments and the
ER's. The post-scission light-particle emission is necessary in order to well
reproduce the measured missing charge distributions obtained in exclusive
fragment-fragment coincidence experiments.

\medskip

However many problems still have to be resolved in order to establish the
systematic behavior of the fusion-fission process in the light-mass region. The
estimate of the barrier height of the scission point has to be more
quantitatively investigated although a linear-dependence of the neck degree of
freedom with the fissility of the compound system seems to be a realistic
approximation. Of course the consistent introduction of deformation effects for
both ER's and complex fragments has to be considered for the systematic
estimation. On the other hand the direct use of OM transmission coefficients
for evaluating the phase-space integration of the complex-fragment binary-decay
is highly desirable to uncover the more interesting features which are included
in the EHFM. Future studies will be undertaken in these directions \cite{Be97}.

\bigskip

\centerline{\bf ACKNOWLEDGEMENTS}

The authors wish to thank R.M. Freeman for fruitful discussions and for a
careful reading of the manuscript. One of us (C.B.) wishes also to acknowledge
S.J. Sanders and Raj K. Gupta for enlightening comments on the model
calculations of Ref.\cite{Sa91,Gu84}. The large part of the EHFM+CASCADE
calculations has been carried out on the HITACHI-3050RX Work Station at Shinshu
University and DEC-osf1 at RIKEN Japan.

%
%

\begin{figure}

Fig.1 : Comparisons of the experimental mass distributions (open
histograms) measured for the $^{32}$S+$^{24}$Mg reaction at E$_{lab}$ = 121
(a) and 142 MeV (b),respectively with the EHFM+CASCADE calculations
(solid histograms). (c) Fragment mass dependence of the lowest scission point
barrier height in the set of the binary combination with same mass number but
different atomic number for $^{56}$Ni. The barrier heights are shown for the
case of angular momentum ${L=0}$. (see text).

\end{figure}

\begin{figure}
Fig.2 : Experimental charge distributions measured for the
$^{35}$Cl+$^{12}$C reaction at E$_{lab}$ = 180, 200 MeV and 278 MeV
and for the $^{37}$Cl+$^{12}$C reaction at E$_{lab}$ = 150 MeV. Comparisons
with EHFM+CASCADE calculations are shown by solid histograms. The dashed
histograms are the results of first-chance EHFM calculations.

\end{figure}

\begin{figure}
Fig.3 : (a) Internal excitation-energy and angular-momentum distributions of
the $^{12}$C fragment as obtained by the first-chance EHFM calculations. The
distributions of the binary-partner nucleus $^{35}$Cl of the $^{12}$C fragment
are shown in (b). (see text).

\end{figure}

\begin{figure}

Fig.4 : Comparisons of the experimental inclusive kinetic energy spectra
measured at the laboratory angle ${\theta_{lab} = 7^o}$ for each fragments with
atomic number from ${Z = 5}$ to ${Z = 11}$ of the $^{35}$Cl+$^{12}$C reaction
at the incident energies E$_{lab}$ = 278 MeV (a) and E$_{lab}$ =
180 MeV (b) with the kinetic energy distributions as obtained with
the EHFM and the EHFM+CASCADE calculations. The solid lines are the kinetic
energy distributions evaluated by including the effects of the post-scission
light-particle CASCADE-decay to the kinetic energy distributions obtained by
the first-chance EHFM calculations shown in dotted lines (see text).

\end{figure}

\begin{figure}
Fig.5 : The missing charge distributions (solid histograms) for the first
fragment with atomic number from ${Z_1 = 5}$ to ${Z_1 = 18}$ as obtained in the
coincidence measurements of the $^{35}$Cl+$^{12}$C reaction at E$_{lab}$ = 278
MeV under the experimental condition of ${Z_1}$ and ${Z_2 \geq 5}$.
The dashed histograms are the calculated values as predicted by EHFM+CASCADE
with the same conditions.

\end{figure}

\begin{figure}
Fig.6 : Comparisons of the mean values ${\langle Z_1+Z_2 \rangle}$ as
obtained by the EHFM+CASCADE calculations (solid lines) with the corresponding
results of the coincidence measurements for the $^{35}$Cl+$^{12}$C reaction at
the incident energies E$_{lab}$ = 200 MeV (a) and 278MeV (b) which are shown as
a function of the first fragment's atomic number ${Z_1}$. The dashed lines
represent the total charge of the compound nucleus Z$_{CN}$ = 23).

\end{figure}

%
%


\begin{references}

\bibitem[*]{Matsuse}
On leave from the Faculty of Textile Science and Technology, Shinshu University,
Ueda, Nagano, 386, Japan, as an Overseas Research Scholar of Japan.

\bibitem{So84} L. G. Sobotka, M. A. McMahan, R. J. McDonald, C. Signarbieux,
G. J. Wozniak, M. L. Padgett, J. H. Gu, Z. H. Liu, Z. Q. Yao and L. G. Moretto,
{\it Phys. Rev. Lett.} {\bf 53}, 2004 (1984).

\bibitem{Sa86} S. J. Sanders, R. R. Betts, I. Ahmad, K. T. Lesko, S. Saini,
B. D. Wilkins, F. Videbaek, B. K. Dichter, {\it Phys. Rev.} {\bf C34}, 1746
(1986).

\bibitem{Sh87} B. Shivakumar, S. Ayik, B. A. Harmon, D. Shapira, {\it Phys.
Rev.} {\bf C35}, 1730 (1987); S. Ayik, D. Shapira, B. Shivakumar, {\it Phys.
Rev.} {\bf C38}, 2610 (1988).

\bibitem{Au87} F. Auger, B. Berthier, A. Cunsolo, A. Foti, W. Mittig, J.M.
Pascaud, E. Plagnol, J. Quebert and J. P. Wieleczko , {\it Phys. Rev.} {\bf
C35}, 190 (1987).

\bibitem{Ch88} R. G. Charity, M. A. McMahan, G. J. Wozniak, R. J. McDonald,
L. G. Moretto, D. G. Sarantites, L. G. Sobotka, G. Guarino, A Pantaleo, L.
Fiore, A. Gobbi and K. D. Hildenbrand, {\it Nucl. Phys.} {\bf A483},371 (1988).

\bibitem{Go91} J. Gomez del Campo, J. L. Charvet, A. D'Onofrio, R. L. Auble,
J. R. Beene, M. L. Halbert,H. J. Kim, {\it Phys. Rev. Lett.} {\bf 61}, 290
(1988); J. Gomez del Campo, R. L. Auble, J. R. Beene, M. L. Halbert,H. J. Kim,
A. D'Onofrio, J. L. Charvet, {\it Phys. Rev.} {\bf C43}, 2689 (1991).

\bibitem{Be92} C. Beck, B. Djerroud, F. Haas, R. M. Freeman, A. Hachem, B.
Heusch, A. Morsad, M. Youlal, Y. Abe, R. Dayras, J. P. Wieleczko, T. Matsuse
and S. M. Lee, {\it Z.Phys.} {\bf A343}, 309 (1992).

\bibitem{Na92} K. Yuasa-Nakagawa, Y. H. Pu, S. C. Jeong, T. Mizota, Y. Futami,
S. M. Lee, T. Nakagawa, B. Heusch, K. Ieki and T. Matsuse, {\it Phys. Lett.}
{\bf B 283}, 185 (1992); Y. Futami, K. Yuasa-Nakagawa, T. Nakagawa, S.M. Lee,
K. Furutaka, K. Matsuda, K. Yoshida, S. C. Jeong, H. Fujihara, T. Mizota, Y.
Honjo, S. Tomita, B. Heusch, K. Ieki, J. Kasagi, W.Q. Shen and T. Matsuse, {\it
Nucl. Phys.} {\bf A 607}, 85 (1996).

\bibitem{Sa87} S. J. Sanders, D. G. Kovar, B. B. Back, C. Beck, B. K. Dichter,
D. J. Henderson, R. V. F. Janssens, J. G. Keller, S. Kaufman, T.-F. Wang, B. D.
Wilkins, F. Videbaek, {\it Phys. Rev. Lett} {\bf 59}, 2856 (1987); S. J.
Sanders, D. K. Kovar, B. B. Back, C. Beck, D. J. Henderson,  R. V. F.
Janssens, T.-F. Wang and B. D. Wilkins. {\it Phys. Rev.} {\bf C40} 2091 (1989).

\bibitem{Ra91} A. Ray, D. Shapira, J. Gomez del Campo, H. J. Kim, C. Beck,
B. Djerroud, B. Heusch, D. Blumenthal and B. Shivakumar, {\it Phys. Rev.}
{\bf C44}, 514 (1991).

\bibitem{Sa91} S.J. Sanders, {\it Phys. Rev.} {\bf C44}, 2676 (1991) and
references therein.

\bibitem{Ma91} T. Matsuse, S. M. Lee and C. Beck, Proceedings of the Symposium
on Heavy-Ion Physics and its Application, Lanzhoo 1990, p.95. Singapore: World
Scientific 1991; T. Matsuse, S. M. Lee, Y. H. Pu, K. Y. Nakagawa, C. Beck and
T. Nakagawa, Proceedings of the International Symposium Towards a Unified
Picture of Nuclear Dynamics, Nikko June 1991, edited by Y. Abe, S .M. Lee and
F. Sakata, p.112 AIP Conference Proceedings ${N^{o}}$ 250 (1992).

\bibitem{Dj92} B. Djerroud, Ph.D. Thesis, Strasbourg University, 1992, Report
No.  CRN/PN 92/32, 1992.

\bibitem{Be93} C. Beck, B. Djerroud, F. Haas, R. M. Freeman, A. Hachem,
B. Heusch, A. Morsad, M. Vuillet-A-Cilles and S. J. Sanders, {\it Phys. Rev.}
{\bf C47}, 2093 (1993).

\bibitem{Be96} C. Beck, D. Mahboub, R. Nouicer, T. Matsuse, B. Djerroud, R. M.
Freeman, F. Haas, A. Hachem, A. Morsad, M. Youlal, S. J. Sanders, R. Dayras, J.
P. Wieleczko, E. Berthoumieux, R. Legrain, E. Pollacco, S1. Cavallaro, E. De
Filippo, G. Lanzan\`o, A. Pagano and M. L. Sperduto, {\it Phys. Rev.} {\bf
C54}, 227 (1996).

\bibitem{Fa96} K. A. Farrar, S. J. Sanders, A. K. Drummer, A. T. Hasan,
F. W. Prosser, B. B. Back, R. R. Betts, M. P. Carpenter, B. Crowell, M. Freer,
D. J. Henderson, R. V. F. Janssens, T. L. Khoo, T. Lauritzen, Y. Liang,
D. Nisius, A. H. Wuosmaa, C. Beck, R. M. Freeman, Sl. Cavallaro, A. Szanto
de Toledo, {\it Phys. Rev.} {\bf C54}, 1249 (1996).

\bibitem{Ma84} T. Matsuse, Proceedings of the International Symposium on Heavy
Ion Fusion Reactions, Tsukuba (Japan), 1984, p.112; S.M.Lee, W.Yokota and
T.Matsuse, Proceedings of the Many Facets of Heavy-Ion Fusion Reactions,
Argonne 1986, Argonne National Laboratory Report No.ANL-PHY-86-1, p.63.

\bibitem{Ha52} W. Hauser and H. Feshbach, {\it Phys. Rev.} {\bf 87}, 366 (1952).

\bibitem{St84} R. G. Stokstad, in {\it Treatise  on Heavy Ion Science}, edited
by A. Bromley (Plenum, New York, 1984), Vol.III.

\bibitem{Mo75} L. G. Moretto, {Nucl. Phys.} {\bf A247}, 211 (1975).

\bibitem{Pu77} F. P\"uhlhofer, {\it Nucl.Phys.} {\bf A280}, 267 (1977).

\bibitem{Ga80} PACE is a modified version of the code JULIAN described by A.
Gavron, {\it Phys. Rev.} {\bf C21}, 230 (1980).

\bibitem{Go81} J. Gomez del Campo and R.G. Stokstad, LILITA, a Monte Carlo
Statistical Model code, ORNL Report TM-7295 (1981).

\bibitem{Aj91} F. Ajzenberg-Selove, {\it Nucl. Phys.} {\bf A 460},1 (1986);
{\bf 475},1 (1987); {\bf 490},1 (1988); {\bf 506}, 1 (1990); {\bf 523},1
(1991).

\bibitem{En90} P. M. Endt, {\it Nucl. Phys.} {\bf 521},1 (1990), and references
therein.

\bibitem{Cst87} D .R. Chakrabarty, S. Sen, M. Thoennessen, N. Alamanos, P. Paul,
R. Schicker, J. Stachel and J. J. Gaardhoje, {\it Phys. Rev.} {\bf C36} 1886
(1987).

\bibitem{Wa85} A. H. Wapstra and G. Audi, {\it Nucl.Phys}. {\bf A432} 1 (1985).

\bibitem{Bm69} A. Bohr and B.R. Mottelson, {\em{Nuclear Structure}} (Benjamin,
New York, 1969), Vol.1.

\bibitem{Fo91} B. Fornal, F. Gramegna, G. Prete, R. Burch, G. D'Erasmo, E. M.
Fiore, A. Pantaleo, V. Paticchio, G. Viesti, P. Blasi, M. Cinausero, F.
Lucarelli, M. Anghinolfi, P. Corvisiero, M. Taiuti, A. Zucchaitti, P. F.
Bortignon, D. Fabris, G. Nebbia and J.A. Ruiz, {\it Phys. Rev. } {\bf C44},
2588 (1991).

\bibitem{Sh91} S. Shlomo and J. B. Natowitz, {\it Phys. Rev. } {\bf C44}, 2878
(1991).

\bibitem{Be85} C. Beck, F. Haas, R.M. Freeman, B. Heusch, J.P. Coffin, G.
Guillaume, F. Rami, and P. Wagner {\it Nucl. Phys.} {\bf A442}, 320 (1985).

\bibitem{Be89a} C. Beck, D.G. Kovar, S. J. Sanders, B. D. Wilkins, D. J.
Henderson, R. V. F. Janssens, W. C. Ma, M. F. Vineyard, T. F. Wang, C. F.
Maguire, F. W. Prosser and G. Rosner {\it Phys. Rev.} {\bf C39}, 2202 (1989).

\bibitem{Be97} C Beck, T. Matsuse and R. Nouicer, in preparation

\bibitem{Hu89} J. R. Huizenga, A. N. Bekhami, I. M. Govil, W. U. Schr\"oder and
J. T\"oke, {\it Phys. Rev. } {\bf C40}, 668 (1989).

\bibitem{Na77} J. R. Natowitz, M. N. Eggers, P. Gonthier, K. Geoffroy, R. Hanus,
C. Towsley and K. Das, {\it Nucl. Phys.} {\bf A277}, 477 (1977).

\bibitem{Va83} N. Van Sen, R. Darves-Blanc, J.C. Gondrand and F. Merchez,
{\it Phys. Rev.} {\bf C27}, 194 (1983).

\bibitem{Pu96} Y. H. Pu, S. M. Lee, S. C. Jeong, H. Fujiwara, T. Mizota, Y.
Futami, T. Nakagawa, H. Ikezoe and Y. Nagame, {\it Z. Phys.} {\bf A353}, 387
(1996).

\bibitem{Bes96} C. Beck and A. Szanto de Toledo, {\it Phys. Rev.} {\bf C53},
1989 (1996).

\bibitem{Kr79} H.J. Krappe, J.R. Nix and A.J. Sierk, {\it Phys. Rev.} {\bf
C20}, 992 (1979).

\bibitem{Du92} D. Durand, {\it Nucl. Phys.} {\bf A541}, 266 (1992).

\bibitem{Go96} J. Gomez del Campo, D. Shapira, M. Korolija, H. J. Kim, K. Teh,
J. Shea, J.P. Wieleczko, E. Ch\'avez, M.E. Ortiz, A. Dacal, C. Volant and A.
D'Onofrio, {\it Phys. Rev.} {\bf C53}, 222 (1996).

\bibitem{Ig75} A.V. Ignatyuk, G.N. Smirenkin and A.S. Tishin, {\it Sov. J. Nucl.
Phys.} {\bf 21}, 255 (1975); A.V. Ignatyuk, K.K. Istekov and G.N. Smirenkin,
{\it Sov. J. Nucl. Phys.} {\it 29}, 250 (1979); O.D. Grudzevich, A.V. Ignatyuk,
V.I. Plyaskin and A.V. Zelenetsky, {\it Proceedings of the International
Conference on Nuclear Data for Science and Technology}, Mito, Japan, 1988
(unpublished).

\bibitem{Ma88} T. Matsuse and S.M. Lee, {\it Proceedings of the International
Conference on Nuclear Data for Science and Technology}, Mito, Japan, 1988,
p.699 (unpublished).

\bibitem{Be89} C. Beck, B. Djerroud, B. Heusch, R. Dayras, R. M. Freeman,
F. Haas, J. P. Wieleczko and M. Youlal, {\it Z. Phys.} {\bf A334}, 521 (1989).

\bibitem{Gu84} R.K. Gupta, M. M\"unchow, A. Sandulescu and W. Scheid, {\it J.
Phys. G: Nucl.Phys.} {\bf 10}, 209 (1984); D.R. Sahora, N. Malhotra and R.K.
Gupta, {\it J. Phys. G: Nucl.Phys.} {\bf 11}, L27 (1985); S.S. Malik and R.K.
Gupta, {\it J. Phys. G: Nucl.Phys.} {\bf 12}, L161 (1986).

\bibitem{Yo89} W. Yokota, T. Nakagawa, M. Ogihara, T. Komatsubara, Y. Fukuchi,
K. Suzuki, W. Galster, Y. Nagashima, K. Furuno, S. M. Lee, T. Mikumo, K. Ideno,
Y. Tomita, H. Ikezoe, Y. Sugiyama and S. Hanashima, {\it Z. Phys.} {\bf A333},
379 (1989).

\bibitem{No96} R. Nouicer, C. Beck, D. Mahboub, T. Matsuse, B. Djerroud, R. M.
Freeman, A. Hachem, Sl. Cavallaro, E. De Filippo, G. Lanzan\`o, A. Pagano and
M. L. Sperduto, R. Dayras, E. Berthoumieux, R. Legrain and E. Pollacco, {\it Z.
Phys.} {\bf A356}, 5 (1996).

\bibitem{Ol96} J. M. Oliveira, Jr., et al., {\it Phys. Rev.} {\bf C53}, 2926
(1996).

\end{references}
\end{document}